# USBee: Air-Gap Covert-Channel via Electromagnetic Emission from USB

Mordechai Guri, Matan Monitz, Yuval Elovici
{gurim,monitzm,elovici}@post.bgu.ac.il
Ben-Gurion University of the Negev
Cyber Security Research Center

*Abstract*— In recent years researchers have demonstrated how attackers could use USB connectors implanted with RF transmitters to exfiltrate data from secure, and even air-gapped, computers (e.g., COTTONMOUTH in the leaked NSA ANT catalog). Such methods require a hardware modification of the USB plug or device, in which a dedicated RF transmitter is embedded.

In this paper we present 'USBee,' a software that can utilize an *unmodified* USB device connected to a computer as a RF transmitter. We demonstrate how a software can intentionally generate controlled electromagnetic emissions from the data bus of a USB connector. We also show that the emitted RF signals can be controlled and modulated with arbitrary binary data. We implement a prototype of USBee, and discuss its design and implementation details including signal generation and modulation. We evaluate the transmitter by building a receiver and demodulator using GNU Radio. Our evaluation shows that USBee can be used for transmitting binary data to a nearby receiver at a bandwidth of 20 to 80 BPS (bytes per second).

## I. INTRODUCTION

Leaking information from a compromised network is one of the main goals of an advanced persistent threat attack. In many cases, common security measures such as firewalls, IDS, and IPS can provide a basic level of protection to secure the internal network and its data. However, when highly sensitive data is involved, the organization may resort to air-gap isolation, where there is no physical connection between the internal network and the Internet.

Over the years, a wide range of covert channels have been proposed to demonstrate how malware can leak data from air-gapped computers without the need for Internet connectivity or physical access. Such covert channels may use electromagnetic, acoustic, thermal, and optical emissions [1] as a medium for data exfiltration from a computer. In 2014, the ANT catalog leaked by Eduard Snowden, present COTTONMOUTH, a tool which allows air-gap communication with a host software, over a USB dongle implanted with an RF transmitter and receiver [2]. Later, in 2015, hackers inspired by COTTONMOUTH introduced TURNIPSCHOOL, a $20 hardware implant concealed in a USB cable which provides short-range RF communication capability to a computer [3]. Hardware based USB keyloggers which include internal radio or Wi-Fi transmitters also exist [4]. However, all of the aforementioned tools require hardware modification of the USB plugs (embedding an RF transmitter or receiver within them). In this paper we show how to leak data from an air-gapped computer over RF signals to a receiver located a short distance away using an *unmodified* USB dongle. We introduce USBee, a malware which utilizes the USB data bus in order to create electromagnetic emissions from a connected USB device. USBee can modulate any binary data over the electromagnetic waves and transmit it to a nearby receiver. The attack scenario is illustrated in Figure 1.

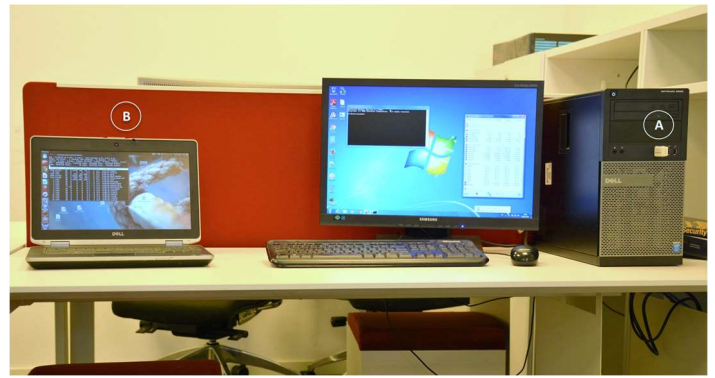

**Figure 1. Illustration of USBee. An ordinary, unmodified USB device (flash drive) (A) is transmitting information to a nearby receiver (B) over an air-gap, via electromagnetic waves emitted from its data bus.**

In this scenario, USBee software, installed on a compromised compute, uses a USB thumb drive already connected to the computer (Figure 1, A), and creates a short-range RF transmission modulated with data (e.g., passwords or encryption keys). The transmission can be received by a nearby receiver (Figure 1, B) where it is decoded and sent to an attacker.

The contribution of our paper is as follows. We introduce a software-only method for short-range data exfiltration using electromagnetic emissions from a USB dongle. Unlike other methods, our method doesn't require any RF transmitting hardware, since it uses the USB's internal data bus. We also discuss signal generation, transmission, reception, and demodulation algorithms.

This paper is organized as follows. Section II presents related work. Section III provides technical background. Section IV and Section V describe transmission and reception. Section VI discusses countermeasures. We conclude with Section VII.

## II. RELATED WORK

Out-of-band covert channels have been discussed since the 1990s. Suggested methods exploit various types of emanation from different computer components in order to modulate and transmit data. There are four covert channel categories: acoustic, optical, thermal, and electromagnetic.

Acoustic methods, discussed in [5] [6] [7] [8], involve transmitting sonic or near-ultrasonic signals from computer speakers. These signals are received and decoded by a microphone in a nearby computer, laptop, or mobile phone. In 2016, Guri et al introduced Fansmitter [9] and DiskFiltration [10], acoustic covert channels from air-gapped computers without speakers or audio hardware. A review of acoustic covert channels is provided in [11]. Optical covert channels have been discussed by Loughry and Umphress [12]. They propose using various computer LEDs to exfiltrate data to a remote camera and implemented a malware which modulates data over the blinking of LEDs in a computer keyboard. VisiSploit [13] is another optical covert channel in which data is leaked from LCD screen to a remote camera via invisible image. An interesting thermal covert channel called BitWhisper [14] demonstrates a communication channel between air-gapped computers that works over thermal pings. Measuring out-of-band covert channels has been discussed in [15]. Over the years, electromagnetic emission has probably been the most researched method of covert communication. Kuhn and Anderson released several publications related to TEMPEST demonstrating that electromagnetic emissions of a desktop computer can be regulated by appropriate software in an offensive manner [16]. Thiele [17] uses a CRT display to emit AM radio signals to a nearby receiver. AirHopper [18] exploits video cards to bridge the air-gap between isolated computers and nearby mobile phones equipped with FM receivers. In this case, a malware on the compromise computer generates carefully crafted radio signals at FM frequencies from the video card of a desktop computer. These signals are picked up and decoded by a mobile phone located up to a few meters from the transmitting computer. In the same manner, GSMem [1] emits electromagnetic signals at cellular frequencies from the RAM bus of a computer. Funthenna [19] uses the GPIO of embedded devices, and Savat [20] uses emission from specific CPU instructions to establish a covert exfiltration channel. Note that compromising emanations from the USB has been studied in [21] [22].

**USB Attacks**

In recent years, several attacks using malicious USB devices have been introduced. Firmware modification of USB controllers was recently made public in the BadUSB research published by SRLabs [23]. In 2014, part of a catalog leaked by Eduard Snowden introduced a tool (COTTONMOUTH) which enabled covert communication using a USB dongle implanted with a RF transmitter and receiver [2]. Later, in 2015, hackers introduced TURNIPSCHOOL, an inexpensive ($20) make-it-yourself hardware hidden in a USB cable which provides RF communication capability to a computer [3]. Several commercial USB based keyloggers which include internal Wi-Fi transmitters also exist [4]. In comparison to the aforementioned methods, our method is software-only and doesn't require firmware or hardware modification.

## III. TECHNICAL BACKGROUND

USB (Universal Serial Bus) is a standard developed in the mid-1990s that defines the equipment (cables and connectors) and communication protocols used in a bus [24]. This definition facilitates communication between computers and peripheral devices. The serial bus is composed of four shielded wires: VBUS, GND, D+, and D-. VBUS is a red/orange wire used for +5v power. GND is a black/blue wire used for grounding. D+ is a white/gold wire, and D- is a green wire; both are used to transmit differential data signals.

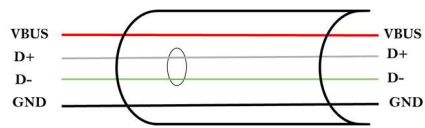

Figure 2. USB wires. D+ and D- form the data bus.

For communication, USB uses a non-return-to-zero inverted (NRZI) encoding scheme. In this scheme, '0' is represented by a transition of the signal level (at clock boundary), while '1' is represented by no transition of the signal level.

USB has three main versions of standards: USB 1.0 (and USB 1.1), USB 2.0, and USB 3.0. USB 1.0 works at a data rate of 1.5 Mbits/s (low speed) to 12 Mbits/s (full speed); USB 2.0 works at a data rate of 480 Mbits/s; and USB 3.0 works at a data rate of up to 5Gbits/s. Our method utilizes the transfer rate of 480 Mbits/s for the transmission, a rate which is supported by USB 2.0 and USB 3.0. In USB NRZI encoding, data is represented in terms of J and K. J means that current is fed into the D+ line, and K means that current is fed into the D- line. For '1' bit, the D+ and D- voltage levels do not change, and they maintain their previous state. For '0' bit, the differential data lines switch their voltage levels (J-K toggle). For example, if D+ is high and D- is low, D+ changes to low and D- changes to high. Bit stuffing is a technique used to achieve clock synchronization by inserting a '0' bit after every six consecutive '1' bits. This way, a sequence of seven '1' bits on the data bus is handled as an error and helps keep the clocks synchronized. There is a side effect related to bit stuffing in that following bit stuffing, the transmission of consecutive '1' bits takes longer than consecutive '0' bits.

The following formula represents the time a data packet takes to transfer (including headers and sync overhead), based on USB 2.0's speed [25].

$$T = (55 * 8 * 2.083) + (2.083 * Floor(3.16 + BitStuffTime(Data\_bc))) + Host\_Delay$$

Where $Data\_bc$ is the data length in bytes, $BitsSuffTime$ is a function that at worst multiplies $Data\_bc$ by $\left(\frac{7}{6}\right)$ for cases in which the data is all '1' and $Host\_delay$ is the time it takes the software components of the mechanism (primarily the controller firmware) to process the transaction.

*A. Electromagnetic Radiation*

Electromagnetic radiation (EMR) is a type of energy emitted by certain electromagnetic processes. EMR propagates through space in the form of electromagnetic waves. These waves have two main properties: (1) frequency $f$ measured in Hertz (Hz), and (2) amplitude measured in decibel-milliwatts (dBm). Electromagnetic waves are generated whenever charged particles are accelerated. Generally speaking, a change of currency in a metal wire creates an electromagnetic emission. The way in which charges and currents interact with the electromagnetic field is described by Maxwell's equations and the Lorentz force law. Electronic devices such as computer monitors, video cards, and cables, emit EMR. The frequency and amplitude depend on the internal current and voltage of the device [26]. Previous research has focused on utilizing EMR for eavesdropping purposes and exploiting the intentional and unintentional EMR of different computer components to create covert channels. In USBee, data is manipulated in the USB's D+ and D- data wires to generate EMR.

## IV. TRANSMISSION

The application layer of the USB protocol provides a definition of the term, endpoint, which can transmit data in only one direction. OUT endpoint can carry data from the host computer to the device (e.g., when writing a file to a USB key). IN endpoint can carry data from the device to the host computer (e.g., when receiving a video feed from a USB webcam). We found that the transmission of a sequence of '0' bits to a USB device generates a detectable emission between $240 Mhz$ and $480 Mhz$. This is understandable given USB 2.0's clock speed and the fact that '0' bits generate a rapid voltage change on each clock cycle as per the USB NRZI implementation. By intentionally sending data from a computer host to a USB device, we can generate controllable EMR that can carry modulated data. A nearby RF receiver can then receive the EMR and decode the data.

*A. Signal Generation*

The basic building block of our transmitter is the generation of electromagnetic signal at a required frequency (Algorithm 1). To that end we created the function, $fill\_buffer\_freq()$. This function receives a pointer to a buffer in the memory (*buf*), the size of this buffer (*size*), and a frequency value (*freq*) which is the target emission frequency in multiples of $100 Khz$. The function fills the buffer with a special pattern which, when transmitted over the data bus (D+ and D-), creates an electromagnetic emission at frequency *freq*. Technically, it fills a buffer in a pattern of bits in such a way that a square wave will be formed. After initializing counter variables $x$ and $i$, we set the variable $t$ as double the ratio between the desired emission frequency and the sample frequency of 480Mbits/s.

```
inline static void fill_buffer_freq
(u32 *buf, int size, double freq)
1    int i = 0;
2    u32 x = 0;
3    double t = freq / 4800 * 2;
4    for (i = 0, x = 0x00000000; i<size*8; i++)
5    {
6        x = x<<1;
7        if ((int)(i*t)%2==0)
8            x++;
9        if((i%32)==31)
10       {
11           *(buf++) = x;
12           x=0x00000000;
13       }
14   }
```
Algorithm 1. The basic signal generation code

Note that since the input frequency is given in multiples of $100 Khz$, it is divided by $4800 Khz$ ($480 Mhz/100$). The main loop (lines 4-14) fills bit elements of the output buffer (*buf*). To build the carrier wave, it fills the array with sequences of '0' and '1' bits which represent the carrier wave at the desired frequency (*freq*). For example, for *freq* = 200, we will obtain a sequence of 4800/100 = 24 bits. This output buffer will be a repeating pattern of 12 ones followed by 12 zeros (111111111111000000000000,..). In line 6 we insert the least significant bit in every iteration. If the expression in line 7 is true, the bit will be set to '1' which based on USB NRZI encoding results imply no change in differential voltage. If the expression is false, the next bit will be flipped, resulting in a differential voltage change when it is transmitted over the bus. Every 32 bits are added to the buffer (line 11). This buffer, when written to a file on a removable storage device or sent to some other USB device, will generate a signal at the desired frequency. The buffer size (*size*) determines the length of the signal. We used a buffer size of 6K, which generates a signal that is strong enough to be clearly detected by the receiver.

*B. Data Modulation*

For digital data modulation we used binary frequency shift keying (B-FSK). Generally, in frequency shift keying data is represented through the frequency changes in a carrier wave, such that distinct frequencies represent distinct digital values. In the binary form of FSK, the frequency of the carrier signal is switched between two values $f_1$ and $f_2$, corresponding to a binary 0 or 1.

*C. Data Transmission*

The actual data transmission is done by writing the byte pattern generated by *fill_buffer_freq* to an arbitrary data block or stream in the USB device. For our purposes, we used a temporary file within the USB thumb drive's file system. The transmission process doesn't require special privileges (e.g., root or admin). It only requires permission to create a file on the removable device. Such files are frequently created by processes for file saving and storing temporary data, or to process intermediate results.

## V. Reception

In order to examine the reception quality, error rate, and effective distance, we built a receiver and demodulator. For a hardware receiver, we used a $30 RTL-SDR software-defined radio connected to a laptop. Our reception code was implemented using GNU Radio [27]. This is a free, open-source software toolkit that provides signal processing building blocks for professional or amateur developers and academic researchers.

### A. Demodulation

As previously described, the transmitted data is modulated using B-FSK. To demodulate the data, we had to convert the received electromagnetic sampled signal from the time domain to the frequency domain using fast Fourier transform (FFT). Note that FFT is a discrete action, whereas the signal is continuous, so we had to cut the signal into chunks before applying FFT. The sampled signal was cut into equal parts and calculated using the following formula:

$$\frac{SR \cdot TOB}{4} = FFTsize$$

Where $SR$ is the sample rate (in $samples/sec$), $TOB$ is the total time it takes to send one bit, and $FFTsize$ is the chunk size. Once the FFT of a chunk has been computed, the resulting vector is passed through an analysis chain. The analysis chain determines the result of the current signal demodulation, which can be either: (1) a '0' bit; (2) a '1' bit, or (3) no signal was detected. Figure 3 presents the FFT results upon receiving a transmitted byte with a value of 0x73 (01110011b). The $X$ axis represents the time in seconds, the $Y$ axis represents the FFT vector index signifying the different frequency levels, and the $Z$ axis represents the amplitude of the frequency component. Low values (noise) were omitted for clarity. As can be seen, the first FFT peak (at 33.07s) has a lower frequency index, meaning this is a '0' bit in the B-FSK modulation. The bit that follows has a higher frequency index, hence representing a '1' bit, and so on. With this modulation and demodulation scheme we successfully transmitted bytes at an effective rate of 80 bytes / sec.

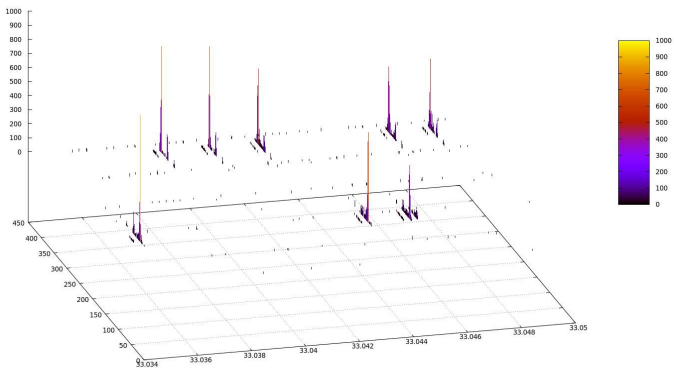

**Figure 3. Reception of 0x73 (01110011b), with FFT over time**

## VI. Countermeasures

Countermeasures can be categorized as procedural, software, and physical. In procedural countermeasures the zones [28] approach may be used. In this approach sensitive computers are kept in restricted areas in which electronic equipment is not allowed. The zones approach, or as it is also known, the 'black-red separation' approach is discussed in [29] to address various types of acoustic, electromagnetic, and optical threats. However, a policy of keeping RF receivers separated or away from computers is not always feasible. The software based approach involves using anti-virus and intrusion detection programs to detect malicious activities. To detect USBee, the I/O operation of a process may be monitored to identify specific patterns (for example, patterns that reflect frequent creation and writes to temporary files). Such behavioral detection mechanisms can suffer from a high rate of false positives, mainly due to the nature of USBee's operations (writing 6K of data to temporary files). Physical isolation involves shielding or preventing EMR from USB components [30]. While grounding and shielding are typically used, another method centers on limiting the amount of emissions generated. USB testing requires scanning the product through frequencies of up to 1 GHz to determine whether any signal amplitudes outside of the specification exist. According to the USB 2.0 specifications, extrapolating EMI by scanning only a few frequencies isn't permitted. It is possible to perform brief testing at the design stage using several antennas that cover different frequency ranges and take into account both horizontal and vertical orientations.

## VII. Conclusion

In this paper we present USBee, a new (software-only) method that turns virtually any USB connector into a short-range RF transmitter. Our method utilizes the data bus in a USB connector to generate electromagnetic radiation of a specific frequency. Code on a contaminated computer can modulate data and transmit it to a nearby receiver, thus creating a type of covert communication channel. Unlike previous covert channels based on USB, our method doesn't require firmware or modification of the USB's hardware. We present the signal generation algorithm and discuss data modulation and demodulation. Our tests show that USBee can be used to effectively transmit data to a nearby receiver at a bandwidth of 80 bytes per second.